\documentclass[conference]{IEEEtran}
\usepackage{cite}
\usepackage{amsmath,amssymb,amsfonts}
\usepackage{algorithmicx}
\usepackage{graphicx}
\usepackage{multirow}
\usepackage{textcomp}
\usepackage{adjustbox}
\usepackage{tikz}
\usetikzlibrary{positioning}
\usepackage{algpseudocode}
\usepackage[ruled,vlined]{algorithm2e}
\usepackage{url}

\begin{document}
\title{Estimating Train Delays in a Large Rail Network Using a Zero Shot Markov Model}


\author{\IEEEauthorblockN{Ramashish Gaurav}
\IEEEauthorblockA{Nutanix Technologies India Pvt Ltd \\
Email: ramashish.gaurav@nutanix.com}
\and
\IEEEauthorblockN{Biplav Srivastava}
\IEEEauthorblockA{IBM T J Watson Research Center, USA\\
Email: biplavs@us.ibm.com}}

\maketitle

\begin{abstract}
India runs the fourth largest railway transport network size carrying over 8 billion passengers per year. 
However, the travel experience of passengers is frequently 
marked by delays, i.e., late arrival of trains at stations, 
causing inconvenience. 
In a first, we study the systemic delays in train arrivals using n-order Markov frameworks and experiment 
with two regression-based models. 
Using train running-status data collected for two years, we report on an efficient algorithm for estimating  delays at railway stations with near accurate results. This work can help railways to manage their resources, while also helping passengers and businesses served by them to efficiently plan their activities.
\end{abstract}
\section{Introduction}

Trains have been a prominent mode of long-distance travel for decades, especially in the countries with a significant land area and large population. India, with a population of  $1.324$ billion people in 2016, has a railway system of network route length of $66,687$ kilometers, with $11,122$ locomotives, $7,216$ stations, that served $8.107$ billion ridership in $2016$ \cite{railstat}. 
%
%
The Indian railway system is fourth largest in the world in terms of network size. However its trains are plagued with endemic delays that can be credited to (a) obsolete technology, e.g., dated rail engines, (b) size, e.g., large network structure and high railway traffic, (c) weather, e.g.,  fog in winter months in north India and rains during summer monsoons countrywide.


In this paper, we take the initial steps in understanding and predicting train delays. Specifically, we focus on the delays of trains, totaling 135, which pass through the busy Mughalsarai station (Station Code: MGS), over a two year period.
We build an 
$N$-Order Markov Late Minutes Prediction Framework ($N$-OMLMPF) which, as we show, predicts near accurate late minutes at the
stations the trains travel to. To the best of our knowledge, this is the first effort to predict train delays for Indian rail network.  The closest prior work is by  Ghosh et al. \cite{ghosh2011statistical} \cite{ghosh} who study the structure and evolution of Indian Railway network, however, they do not estimate delays. Our analysis is complementary and agrees with the characteristics of the busiest train stations that they find.
We now define the problem, outline contributions, and present our approach. 


\textbf{Problem Statement}: \textit{Given a train and its route information, predict the delay in minutes at an in-line station during its journey on a valid date}.

\subsection{Contributions}
Our main contributions are that we:
\begin{itemize}
\item as a first, present the dataset of 135 Indian trains' running status information (which captures delays along stations), collected for two years. We plan to make it public.

\item build a scalable, train-agnostic, and Zero-Shot competent framework for predicting train arrival delays, learning from a fixed set of trains and transferring the knowledge to an unknown set of trains. 
\item study delays using $n$-order Markov Process Regression models and do Akaike Information Criterion (AIC) and Schwartz Bayesian Information Criterion (BIC) analysis to find the correct order of the Markov Process. Most of the 135 trains follow 1-order Markovian Process.
\item discuss how the train-agnostic framework can leverage different types of trained models and be deployed in real time 
to predict the late minutes at an in-line station. 
\end{itemize}


%
%

The rest of paper is arranged as follows. We first discuss the data about train operation and its analysis in Section \ref{aod} and then present the proposed model in Section \ref{pm}. Next, in section \ref{ra}, we outline the experiments conducted with two different regression models:  Random Forest Regression and Ridge Regression and give an exhaustive analysis of our results. Finally, we conclude with pointers for future research.

\section{Data Preprocessing and Analysis} \label{aod}

This section gives details of train information we collected for a span of 
two years from site\cite{railapi}. Table~\ref{data_descr} gives the statistics.

\begin{table}[b]
\vspace{-10pt}
\centering
\caption{Data Statistics for 135 Trains Complete Data}
\label{data_descr}
\begin{tabular}{|l|c|}
\hline
Total number of trains considered             & 135 \\ \hline
Total number of unique stations covered       & 819 \\ \hline
Maximum number of journeys made by a train    & 334 \\ \hline
Average number of journeys made by a train    & 48  \\ \hline
Maximum number of stations in a train's route & 129 \\ \hline
Average number of stations in a train's route & 30  \\ \hline
\end{tabular}
\end{table}

\subsection{Data Collection and Segregation}

%
%
We considered 135 trains that pass through Mughalsarai Station (MGS), one of top busiest stations in India. For them, we collected train running status information (\textit{Train Data}) over the period of March 2016 to February 2018. A train's \textit{Train Data} consists of multiple instances of journeys, where each journey has the same set of in-line stations that the train plies through. 
Table \ref{tab_rddtj} has important fields of interest in \textit{Train Data}. 

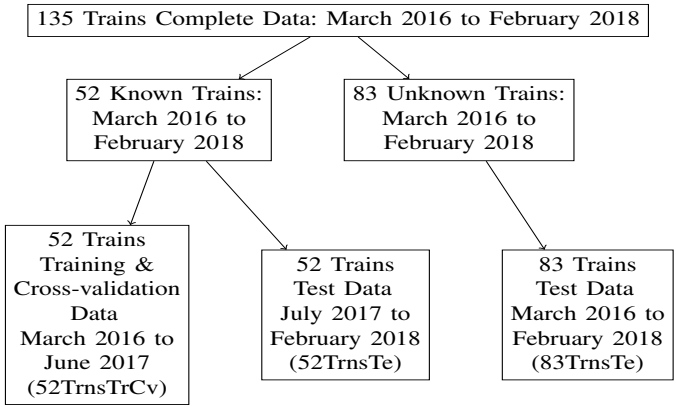
\begin{figure}
\begin{adjustbox}{width=3.5in, totalheight=2.1in}
\begin{tikzpicture}

\node(A)[draw, align=center] at (0.35,0) {135 Trains Complete Data: March 2016 to February 2018};
\node (A4) [draw, below=of A, align=center] at (-2,0) { 52 Known Trains: \\ March 2016 to \\ February 2018};
\node (A5) [draw, right=of A4, align=center] {83 Unknown Trains: \\ March 2016 to \\ February 2018};
\node (A2) [draw, below=of A4, align=center] at (-3,-2.5) {52 Trains \\ Training \& \\ Cross-validation \\ Data\: \\ March 2016 to \\ June 2017 \\ (52TrnsTrCv)};
\node (A1) [draw, right=of A2, align=center] {52 Trains \\ Test Data\: \\ July 2017 to \\ February 2018 \\ (52TrnsTe)};
\node (A3) [draw, right=of A1, align=center]{83 Trains \\ Test Data\: \\ March 2016 to \\ February 2018 \\ (83TrnsTe)};

\draw[->] (A) -- (A4);
\draw[->] (A) -- (A5);
\draw[->] (A4) -- (A2);
\draw[->] (A4) -- (A1);
\draw[->] (A5) -- (A3);
\end{tikzpicture}
\end{adjustbox}
\caption{Segregation of Complete Data of 135 Trains for Experimentation. \newline The complete data is divided into two sets: 52 \textit{Known Trains} and 83 \textit{Unknown Trains}. \textit{Known Trains} data is further subdivided into 52 Trains Training \& Cross-validation Data (52TrnsTrCv) and 52 Trains Test Data (52TrnsTe) with different time periods. The \textit{Unknown Trains} data (83TrnsTe) is kept intact to assess knowledge transfer from \textit{Known Trains} to \textit{Unknown Trains}.}
\label{pict_rep_data}
\vspace{-15pt}
\end{figure}

\begin{figure*}[htbp]
\centering
\begin{adjustbox}{width=\textwidth}
\begin{tikzpicture} [node distance=0.3cm]
\node(A)[draw, fill=green] at (0, 0){\footnotesize RNC};
\node(B)[draw] at (2, 0) {\footnotesize BKSC};
\node(C)[draw] at (4, 0){\footnotesize  KQR};
\node(D)[draw] at (6, 0){\footnotesize GAYA};
\node(E)[draw] at (8, 0){\footnotesize  DOS};
\node(F)[draw] at (10, 0){\footnotesize  MGS};
\node(G)[draw] at (12, 0){\footnotesize CNB};
\node(H)[draw, fill=red] at (14, 0){\footnotesize NDLS};

\draw[->, double distance=2pt] (A) -- (B);
\draw[->, double distance=2pt] (B) -- (C);
\draw[->, double distance=2pt] (C) -- (D);
\draw[->, double distance=2pt] (D) -- (E);
\draw[->, double distance=2pt] (E) -- (F);
\draw[->, double distance=2pt] (F) -- (G);
\draw[->, double distance=2pt] (G) -- (H);

\node (A1) [below=of A, align=center]{ \footnotesize Source \\ \footnotesize  Station};
\node (H1) [below=of H, align=center]{\footnotesize  Destination \\ \footnotesize Station};
\node (F1) [below=of F, align=center]{\footnotesize  Current \\ \footnotesize Station ($Stn_0$)};
\node (E1) [below=of E, align=center]{\footnotesize  $1^{st}$ Previous \\ \footnotesize Station ($Stn_1$)};
\node (D1) [below=of D, align=center]{\footnotesize $2^{nd}$ Previous \\ \footnotesize Station ($Stn_2$)};
\node (C1) [below=of C, align=center]{ \footnotesize $3^{rd}$ Previous \\ \footnotesize Station ($Stn_3$)};
\node (B1) [below=of B, align=center] { \footnotesize $4^{th}$ Previous \\ \footnotesize Station ($Stn_4$)};

\draw[->] (A) -- (A1);
\draw[->] (B) -- (B1);
\draw[->] (C) -- (C1);
\draw[->] (D) -- (D1);
\draw[->] (E) -- (E1);
\draw[->] (F) -- (F1);
\draw[->] (H) -- (H1);
\end{tikzpicture}
\end{adjustbox}
\caption{Train Route of Train 12439. The above figure shows the route of train 12439 which starts at the station $RNC$ and ends at the station $NDLS$. For current station $MGS$, 4 previous stations are considered; whose information we can use for preparing a $4$-$prev$-$stn$ data-frame (Table \ref{my-label}). $Stn_i$ notation for $i^{th}$ previous station is used throughout this paper.}
\label{tr}
\end{figure*}
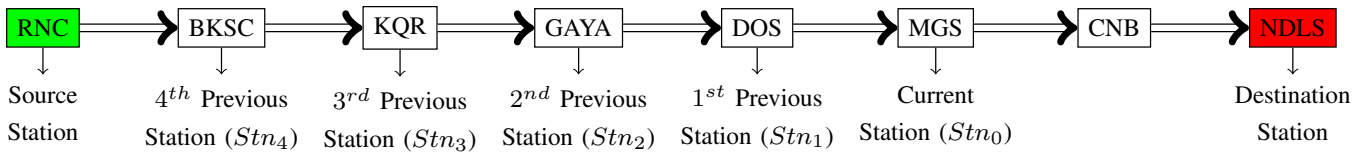

Due to the infrequent running of trains, the amount of data collected for each of the trains greatly varied. 
Using the file size as criterion, we selected \textit{Train Data} of 52 frequent trains (henceforth mentioned as \textit{Known Trains}), out of 135, as training data. 
The data of remaining 83 trains (henceforth mentioned as \textit{Unknown Trains}) were used for testing and evaluating the transfer of knowledge 
through trained models. 
Figure \ref{pict_rep_data} pictorially illustrates the actual segregation of collected \textit{Train Data} from March 2016 to February 2018 for 135 trains. 
One may recall that in traditional machine learning, the training and test data are drawn from the same set (or class). In contrast, 
 we train our models on a seen set of \textit{Known Trains} and test it on an unseen set of \textit{Unknown Trains}, thus employing zero data of \textit{Unknown Trains} for training, hence the term Zero-Shot. 
This problem setting is similar to Zero Shot Learning \cite{zsl} where training and test set classes' data are disjoint. 
Figure \ref{tr} shows a train journey and related notations used in this paper.
\begin{table}[b]

\vspace{-20pt}
\setlength\tabcolsep{3.95pt}
\centering
\caption{Description of \textit{Train Data} collected for each train}
\label{tab_rddtj}
\begin{tabular}{|l|l|}
\cline{1-2}
 \textbf{Field Name} & \textbf{Description} \\ \cline{1-2}
 $actarr\_date$ & Actual arrival date of train at a station e.g. $19$ $Sep$ $2016$  \\ \cline{1-2}
 $station\_code$ & Station code name (acronym) for a station e.g. $MGS$ \\ \cline{1-2}
 $latemin$ & Late minutes (arrival delay) at station e.g. $107$ \\ \cline{1-2}
 $distance$ & Distance of a station from the source in kilometers e.g. $204$ \\ \cline{1-2}
 $month$ & Jan, Feb, Mar... extracted from $actarr\_date$ \\ \cline{1-2}
 $weekday$ & Mon, Tue, Wed... extracted from $actarr\_date$ \\ \cline{1-2}
\end{tabular}

\end{table}

\vspace{-3pt}
\subsection{Data Preparation}\label{ddp} 
We define a \textit{data-frame} as a collection of multiple rows with fixed number of columns.
For our experiments we prepared two types of data-frames, with one type being  
a data-frame Table \ref{my-label} for each station (henceforth mentioned as \textit{Known Stations}, totaling 621 out of 819) 
falling in the journey route of \textit{Known Trains} by extracting required information from \textit{Train Data} Table \ref{tab_rddtj} of respective trains (in whose route the station fell) to train the models. 
Another type consisted of only one data-frame Table \ref{my-label1} capturing certain information of all 819 stations; irrespective of whether they are in-line to \textit{Known Trains} or \textit{Unknown Trains}. We divided the journey data in \textit{52TrnsTrCv} Data in ratio 4 to 1 to train and cross-validate the models and prepared data-frame (Table \ref{my-label}) for the chosen 80\% journey data. However we did not prepare any data-frames (Table \ref{my-label}) for rest 20\% of \textit{52TrnsTrCv} Data, \textit{52TrnsTe} Data and \textit{83TrnsTe} Data, thereby leaving them in their native format of \textit{Train Data} Table \ref{tab_rddtj}. 


\vspace{-5pt}
\subsection{Data Analysis}
\label{da}

\begin{figure}[t]
\vspace{-20pt}
  \centering
  \includegraphics[width=1\linewidth]{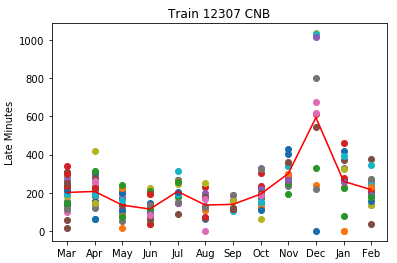}
  \caption{Monthly variation of late minutes at station $CNB$ for Train 12307}
  \label{12307}
\vspace{-15pt}
\end{figure}

\begin{figure}[t]
\vspace{-20pt}
  \centering
  \includegraphics[width=1\linewidth]{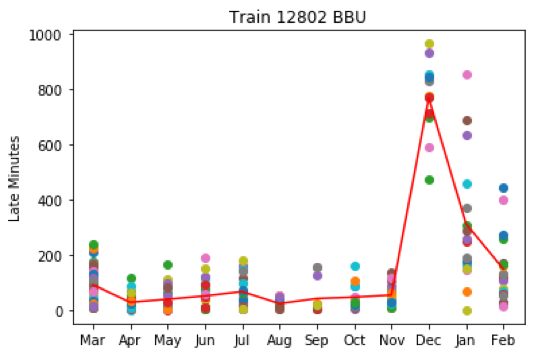}
  \caption{Monthly variation of late minutes at station $BBU$ for Train 12802}
  \label{Picture10}
\vspace{-20pt}
\end{figure}

\begin{figure}[t]

  \centering
  \includegraphics[width=1\linewidth]{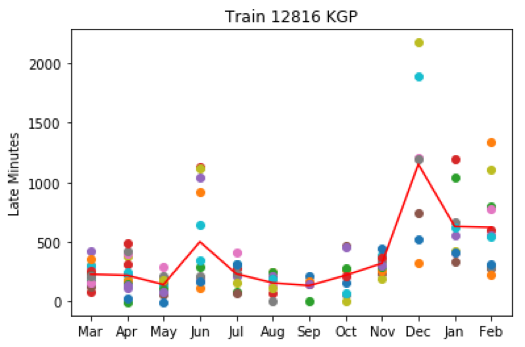}
  \caption{Monthly variation of late minutes at station $KGP$ for Train 12816}
  \label{Picture5}
\vspace{-10pt}
\end{figure}

\begin{figure}[t]
  \centering
  \includegraphics[width=1\linewidth]{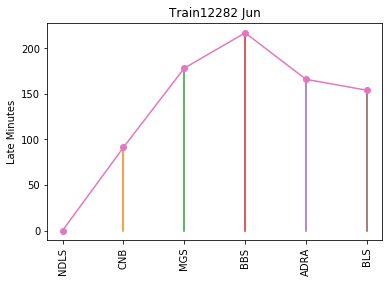}
  \caption{Mean late minutes during Train 12282's journey in June 2017}
  \label{12282}
\vspace{-10pt}
\end{figure}

\begin{figure}[t]

  \centering
  \includegraphics[width=1\linewidth]{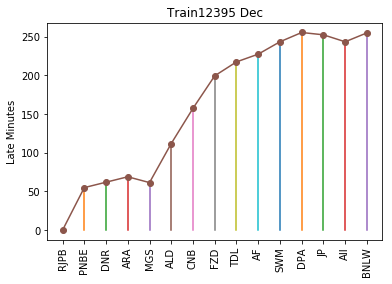}
  \caption{Mean late minutes during Train 12395's journey in December 2017}
  \label{Picture1_t}
\vspace{-10pt}
\end{figure}

\begin{figure}[t]

  \centering
  \includegraphics[width=1\linewidth]{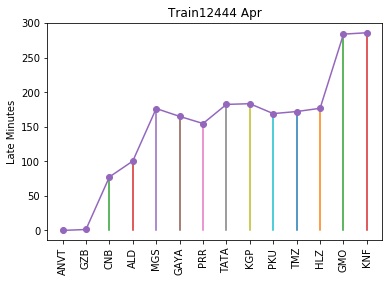}
  \caption{Mean late minutes during Train 12444's journey in April 2017}
  \label{Picture7_t}
\vspace{-15pt}
\end{figure}

Here we analyze the most important factors which drive our learning and prediction algorithm. As observed in Figures \ref{12307}, \ref{Picture10}, and \ref{Picture5}, the spikes in each month signify that mean late minutes at a station varies monthly (the colored dots are the individual late minutes during the month). This premise was verified with similar graphs obtained for other trains and their in-line stations. In Figures \ref{12282}, \ref{Picture1_t}, and  \ref{Picture7_t}, the dots represent the mean of late minutes at each in-line station during a train's journey in a particular month. In Figure \ref{12282} we can see that the mean late minutes increase during journey up-till station $BBS$ and later it decreases. We observed similar graphs for other trains and found that partial sequences of consecutive in-line stations characterize the delays during a train's journey. 

\begin{table*}[!t]
\vspace{-10pt}
\centering
\caption{Description of Training Data prepared from \textit{Train Data} Table \ref{tab_rddtj}}
\label{my-label}
\begin{tabular}{ccccccccc}
\hline
\multicolumn{2}{|c|}{\textbf{train\_type}}                                                                                            & \multicolumn{1}{c|}{\textbf{zone}}                                                               & \multicolumn{2}{c|}{\textbf{is\_superfast}}                                                                                  & \multicolumn{2}{c|}{\textbf{month}}                                                                                                                                                     & \multicolumn{2}{c|}{\textbf{weekday}}                                                                                               \\ \hline
\multicolumn{2}{|c|}{Is it Special, Express or Other?}                                                                                & \multicolumn{1}{c|}{\begin{tabular}[c]{@{}c@{}}What zone does the \\ train belong to?\end{tabular}}   & \multicolumn{2}{c|}{Is it super fast?}                                                                                      & \multicolumn{2}{c|}{Month in which the journey is made}                                                                                                                                 & \multicolumn{2}{c|}{\begin{tabular}[c]{@{}c@{}}Weekday on which\\ the journey is made\end{tabular}}                                 \\ \hline
\multicolumn{5}{|c|}{Obtained from \cite{railfaq} through train number (e.g. 13050 for Train 13050)}                                                                                                                                                                                                                                                                                                       & \multicolumn{4}{c|}{Obtained from \textit{actarr\_date} (Table \ref{tab_rddtj})}                                                                                                                                                                                                                                                                                \\ \hline
\multicolumn{1}{l}{}                                                                              & \multicolumn{1}{l}{}              & \multicolumn{1}{l}{}                                                                             & \multicolumn{1}{l}{}                                                                    & \multicolumn{1}{l}{}              & \multicolumn{1}{l}{}                                                                    & \multicolumn{1}{l}{}                                                                          & \multicolumn{1}{l}{}              & \multicolumn{1}{l}{}                                                                            \\ \hline
\multicolumn{1}{|c|}{\textbf{Stn\textsubscript{1}\_code}}                                                          & \multicolumn{1}{c|}{\textbf{...}} & \multicolumn{1}{c|}{\textbf{Stn\textsubscript{n}\_code}}                                                          & \multicolumn{1}{c|}{\textbf{late\_mins\_Stn\textsubscript{1}}}                                            & \multicolumn{1}{c|}{\textbf{...}} & \multicolumn{1}{c|}{\textbf{late\_mins\_Stn\textsubscript{n}}}                                            & \multicolumn{1}{c|}{\textbf{db\_Stn\textsubscript{0}\_Stn\textsubscript{1}}}                                                    & \multicolumn{1}{c|}{\textbf{...}} & \multicolumn{1}{c|}{\textbf{db\_Stn\textsubscript{n-1}\_Stn\textsubscript{n}}}                                                    \\ \hline
\multicolumn{1}{|c|}{\begin{tabular}[c]{@{}c@{}}Station Code\\ of $Stn_1$\end{tabular}}              & \multicolumn{1}{c|}{...}          & \multicolumn{1}{c|}{\begin{tabular}[c]{@{}c@{}}Station Code\\ of $Stn_n$\end{tabular}}              & \multicolumn{1}{c|}{\begin{tabular}[c]{@{}c@{}}Late Minutes\\ at $Stn_1$\end{tabular}}     & \multicolumn{1}{c|}{...}          & \multicolumn{1}{c|}{\begin{tabular}[c]{@{}c@{}}Late Minutes\\ at $Stn_n$\end{tabular}}     & \multicolumn{1}{c|}{\begin{tabular}[c]{@{}c@{}}Distance between\\ $Stn_0$ and $Stn_1$\end{tabular}} & \multicolumn{1}{c|}{...}          & \multicolumn{1}{c|}{\begin{tabular}[c]{@{}c@{}}Distance between\\ $Stn_{n\!^{\_}1}$ and $Stn_n$\end{tabular}} \\ \hline
\multicolumn{3}{|c|}{Obtained from \textit{station\_code} (Table \ref{tab_rddtj})}                                                                                                                                                                                         & \multicolumn{3}{c|}{Obtained from \textit{latemin} (Table \ref{tab_rddtj})}                                                                                                                                                                            & \multicolumn{3}{c|}{Obtained from \textit{distance} (Table \ref{tab_rddtj})}                                                                                                                                                                                         \\ \hline
\multicolumn{1}{l}{}                                                                              & \multicolumn{1}{l}{}              & \multicolumn{1}{l}{}                                                                             & \multicolumn{1}{l}{}                                                                    & \multicolumn{1}{l}{}              & \multicolumn{1}{l}{}                                                                    & \multicolumn{1}{l}{}                                                                          & \multicolumn{1}{l}{}              & \multicolumn{1}{l}{}                                                                            \\ \hline
\multicolumn{1}{|c|}{\textbf{Stn\textsubscript{1}\_dfs}}                                                           & \multicolumn{1}{c|}{\textbf{...}} & \multicolumn{1}{c|}{\textbf{Stn\textsubscript{n}\_dfs}}                                                           & \multicolumn{1}{c|}{\textbf{tfc\_of\_Stn\textsubscript{1}}}                                               & \multicolumn{1}{c|}{\textbf{...}} & \multicolumn{1}{c|}{\textbf{tfc\_of\_Stn\textsubscript{n}}}                                               & \multicolumn{1}{c|}{\textbf{deg\_of\_Stn\textsubscript{1}}}                                                     & \multicolumn{1}{c|}{\textbf{...}} & \multicolumn{1}{c|}{\textbf{deg\_of\_Stn\textsubscript{n}}}                                                       \\ \hline
\multicolumn{1}{|c|}{\begin{tabular}[c]{@{}c@{}}$Stn_1$ distance\\ from source station\end{tabular}} & \multicolumn{1}{c|}{...}          & \multicolumn{1}{c|}{\begin{tabular}[c]{@{}c@{}}$Stn_n$ distance\\ from source station\end{tabular}} & \multicolumn{1}{c|}{\begin{tabular}[c]{@{}c@{}}Traffic Strength\\ of $Stn_1$\end{tabular}} & \multicolumn{1}{c|}{...}          & \multicolumn{1}{c|}{\begin{tabular}[c]{@{}c@{}}Traffic Strength\\ of $Stn_n$\end{tabular}} & \multicolumn{1}{c|}{\begin{tabular}[c]{@{}c@{}}Degree Strength\\ of $Stn_1$\end{tabular}}        & \multicolumn{1}{c|}{...}          & \multicolumn{1}{c|}{\begin{tabular}[c]{@{}c@{}}Degree Strength\\ of $Stn_n$\end{tabular}}          \\ \hline
\multicolumn{3}{|c|}{Obtained from \textit{distance} (Table \ref{tab_rddtj})}                                                                                                                                                                                             & \multicolumn{6}{c|}{Obtained from Open Government Data (OGD) {[}4{]}}                                                                                                                                                                                                                                                                                                                                                                                       \\ \hline
\multicolumn{1}{l}{}                                                                              & \multicolumn{1}{l}{}              & \multicolumn{1}{l}{}                                                                             & \multicolumn{1}{l}{}                                                                    & \multicolumn{1}{l}{}              & \multicolumn{1}{l}{}                                                                    & \multicolumn{1}{l}{}                                                                          & \multicolumn{1}{l}{}              & \multicolumn{1}{l}{}                                                                            \\ \hline
\multicolumn{2}{|c|}{\textbf{Stn\textsubscript{0}\_dfs}}                                                                                               & \multicolumn{1}{c|}{\textbf{Stn\textsubscript{0}\_tfc}}                                                           & \multicolumn{2}{c|}{\textbf{Stn\textsubscript{0}\_deg}}                                                                                      & \multicolumn{4}{c|}{\textbf{Stn\textsubscript{0}\_late\_minutes}}                                                                                                                                                                                                                                                                               \\ \hline
\multicolumn{2}{|c|}{$Stn_0$ distance from source station}                                                                               & \multicolumn{1}{c|}{$Stn_0$ traffic strength}                                                       & \multicolumn{2}{c|}{$Stn_0$ degree strength}                                                                                   & \multicolumn{4}{c|}{Current Station's target late minutes to be predicted}                                                                                                                                                                                                                                                    \\ \hline
\multicolumn{2}{|c|}{Obtained from \textit{distance} (Table \ref{tab_rddtj})}                                                                                          & \multicolumn{3}{c|}{Obtained from OGD {[}4{]}}                                                                                                                                                                                 & \multicolumn{4}{c|}{Obtained from \textit{latemin} (Table \ref{tab_rddtj})}                                                                                                                                                                                                                                                                                    \\ \hline
\end{tabular}
\\~\\ The bold font texts are the columns in our prepared data-frame for each \textit{Known Station}. We assert that \textbf{Stn\textsubscript{0}\_late\_minutes} depends on the values mentioned in other columns. \textbf{tfc\_of\_Stn\textsubscript{i}} and \textbf{deg\_of\_Stn\textsubscript{i}} are the total number of trains passing through $Stn_i$ and total number of direct connections of $Stn_i$ to other stations respectively. Such a data-frame is called $n$-$prev$-$stn$ data-frame of a target station ($Stn_0$) for which it is prepared, where $n$ depends on the number of previous stations (a partial sequence of consecutive stations) considered.
\end{table*}

\begin{table}[]
\vspace{-18pt}
\centering
\caption{Description of \textit{Station Features}}
\label{my-label1}
\begin{tabular}{|c|c|c|c|c|}
\hline
\textbf{station}                                                 & \textbf{latitude}                             & \textbf{longitude}                             & \textbf{stn\_tfc}                                                     & \textbf{stn\_deg}                                                     \\ \hline
$Stn$                                                           & Latitude                                      & Longitude                                      & \begin{tabular}[c]{@{}c@{}}Traffic Strength\\ of station\end{tabular} & \begin{tabular}[c]{@{}c@{}}Degree \\ Strength \\ of station\end{tabular} \\ \hline
\begin{tabular}[c]{@{}c@{}}Obtained from \\ $station\_code$\end{tabular} & \multicolumn{2}{c|}{\begin{tabular}[c]{@{}c@{}}Obtained from \\ Google Maps APIs\end{tabular}} & \multicolumn{2}{c|}{Obtained from OGD \cite{railtt}}                                                                                                        \\ \hline
\end{tabular}
\\~\\The bold font texts are the columns in our prepared data-frame for collectively all 819 stations of \textit{Known Trains} and \textit{Unknown Trains}. \textbf{station} is used as a \textit{key} to obtain rest $4$ features on which k-NN is run. This data- frame helps to determine the semantically nearest station to a given station.  
\end{table}







\vspace{-1pt}
\section{Proposed Model} \label{pm}

In this section, we explain our proposed regression-based $N$-OMLMPF algorithm and its components. Regression is the task of analyzing the effects of independent variables (in a multi-variate data) on a dependent continuous  variable and predicting it. In our setting, the independent variables are the ones mentioned in Table \ref{my-label} and the dependent continuous variable to be predicted is the target late minutes (\textbf{Stn\textsubscript{0}\_late\_minutes}). Our regression experiments with low RMSE and significant accuracy under 95\% Confidence Interval back our hypothesis to cast it as a Regression based problem. We used Random Forest Regressors (RFRs) and Ridge Regressors (RRs) as two types of individual regression models in $N$-OMLMPF to learn, predict, evaluate, and compare results.




For real-time deployment and scalability, we avoided building train-specific models. Hence we looked for entities which would help us to frame a train-agnostic algorithm as well as enable knowledge transfer from \textit{Known  Trains} to \textit{Unknown Trains}. 
A train's route is composed and characterized by the \textit{Stations} in-line in its journey. Significant delays along a route which has more number of busy stations can be expected compared to the ones having lesser number of busy stations. 

Through the analysis of multiple figures similar to the ones mentioned in subsection \ref{da} we observed the following \textit{details} about the delay at in-line stations during a journey:
\begin{itemize}
\item It highly depends on the months during which the journey is made. One can observe the variations during summer ($Jun$ in \figurename  \ref{12307}) and winter months ($Dec$ in \figurename \ref{12307}).
\item Partial routes of consecutive \textit{Stations} can be identified during journey which either increase or decrease the delay at next stations ($CNB\rightarrow MGS\rightarrow BBS$ in \figurename \ref{12282}).
\item \textit{Stations} with a high traffic and degree strength tend to be the bottleneck in a journey, thus increasing the overall lateness ($MGS$-a busy station in \figurename \ref{12282}, \figurename \ref{Picture1_t}, and \figurename \ref{Picture7_t}).
\end{itemize} 

Above points suggest that multiple deciding factors (e.g. the month of travel, the sequence of stations during a journey etc.) determine the late minutes at a station considered. Since we sought to use \textit{Stations} to frame a train-agnostic late minutes prediction algorithm and for knowledge transfer, we prepare a data-frame Table \ref{my-label} for each of the \textit{Known Stations} capturing the \textit{details} mentioned. Later, we train $n$-Order Markov Process Regression models for each \textit{Known Station}; described next.
\subsection{$n$-Order Markov Process Regression ($n$-OMPR) Models} 
The Markov Process asserts that the outcome at a current state depends only on the outcome of the immediately previous state. However if the current state's outcome depends on $n$ previous states, we call it an $n$-Order Markov Process. Here we assert that the late minutes at a current target station depends on  the details of its $n$-previous stations (henceforth mentioned as $n$-$prev$-$stns$). 
This notion is effectively captured in data-frame Table \ref{my-label} where we capture general features of a train, day and month of a journey and the characteristics of the $n$-$prev$-$stns$ along with that of the current target station.  
The idea is to learn $n$-OMPR models (Random Forest Regressors and Ridge Regressors) for each of the \textit{Known Stations} using Algorithm \ref{algo_training} and later use those trained models to frame a train-agnostic late minutes prediction algorithm ($N$-OMLMPF Algorithm \ref{algo_lmsp}). 
Regression models are trained on each of the \textit{Known Stations}' corresponding $n$-$prev$-$stn$ data-frame Table \ref{my-label} with the values of $n$ depending on the number of stations previous to it, subject to its positions during the journeys of multiple trains. This design will be clarified in section \ref{exm1}. We used python sklearn.ensemble library \cite{scikit-learn} and sklearn.linear\_model library \cite{scikit-learn} for learning Random Forest Regressor and Ridge Regressor models respectively. 

\subsection{$k$-Nearest Neighbor ($k$-NN) Search}
\textit{Unknown Stations} (\textit{USs}) are the ones which, along with the \textit{Known Stations} (\textit{KSs}), build the journey route of \textit{Unknown Trains}. Since we made \textit{Unknown Trains}' data Zero Shot, data-frame Table \ref{my-label} is not prepared for \textit{USs}, thus we do not have $n$-OMPR models for them. Hence, we look for a $KS$ which is best similar to the current target $US$ with respect to features stated in Table \ref{my-label1}; whose model could be used to approximate the predicted late minutes at the $US$.
We employ $k$-NN search algorithm (Algorithm \ref{algo_knn}) to fulfill this objective. A two-step $k$-NN search is applied since latitude and longitude data are semantically different from traffic and degree strength data. We used python sklearn.neighbors library \cite{scikit-learn} with default options.

\begin{algorithm}[!h] 
\SetAlgoLined
\small
\caption{Training n-OMPR Models}
\label{algo_training}
\KwIn{List Of \textit{Known Stations} ($KS$): $<KS_1, ... KS_M>$}
\KwOut{$n$-OMPR Models for \textit{Known Stations}}
\For{$i=1; i<=5; i+=1$} {
	$ips_{list}$ $\gets$ Initialize empty list (stores stations having $i$-OMPR models)\\
}
\For{$\forall$ $stn_{k}$ $\in$ \textit{Known Stations}} {
	\For{$i=1; i<=5; i+=1$} {
		$df$ $\gets$ Get $stn_{k}$'s $i$-$prev$-$stn$ data-frame (Table \ref{my-label})\\
        \If{$df$ is not empty} { 
        	$mdl_{i}^{stn_k}$ $\gets$ Train RFR \& RR Models on $df$ \\
            $ips_{list}$ $\gets$ $ips_{list}$ + $stn_{k}$ \Comment{Include $stn_{k}$ in list}\\
            Save $mdl_{i}^{stn_k}$
        }
    }
}
\For{$i=1; i<=5; i+=1$} {
	Save $ips_{list}$
}
\end{algorithm}

\begin{algorithm}[!h]
\SetAlgoLined
\small
\caption{$N$-OMLMPF for \textit{Known Trains} and \textit{Unknown Trains} (here the value of $N$ is set as $3$ $\implies$ limit the models up to $3$-OMPR models)}
\label{algo_lmsp}
\KwIn{Train number $tr_{num}$, in-line stations list ($stn_{jrny}$), journey route information (Table \ref{tab_rddtj}), $ips_{list}$}
\KwOut{A list ($lms_{stn}$) of predicted late minutes at each station during the journey}
$lms_{stn}$ $\gets$ Initialize late minutes list with entry $<0>$ ($0$ minutes late at source)\\
\For{$i=1; i<length(stn_{jrny}); i+=1$} {
	$crnt_{stn}$ = $stn_{jrny}$.At($i$) \Comment{Station at $i^{th}$ position}\\
	\uIf{$crnt_{stn}$ is at position i = 1} {
    	$df_{stn}$ $\gets$ Prepare $crnt_{stn}$'s $1$-$prev$-$stn$ row data-frame (Table \ref{my-label}) using Table \ref{tab_rddtj} with \textbf{late\_mins\_Stn\textsubscript{1}} set as $lms_{stn}$.At(0)\\
        \If{$crnt_{stn}$ $\notin$ $1ps_{list}$} { 
            $crnt_{stn}$ $\gets$ Get nearest \textit{Known Station} in $1ps_{list}$ using Algorithm \ref{algo_knn}
        }
        $lms_{stn}$.At($i$) $\gets$ Predict late minutes at $crnt_{stn}$ for $df_{stn}$ using $mdl_{1}^{crnt_{stn}}$ model\\
    }
    \uElseIf{$crnt_{stn}$ is at position i = 2} {
    	$df_{stn}$ $\gets$ Prepare $crnt_{stn}$'s $2$-$prev$-$stn$ row data-frame (Table \ref{my-label}) using Table \ref{tab_rddtj} with \textbf{late\_mins\_Stn\textsubscript{1}} set as $lms_{stn}$.At(1) and \textbf{late\_mins\_Stn\textsubscript{2}} set as $lms_{stn}$.At(0) \\
        \If{$crnt_{stn}$ $\notin$ $2ps_{list}$}{ 
			$crnt_{stn}$ $\gets$ Get nearest \textit{Known Station} in $2ps_{list}$ using Algorithm \ref{algo_knn}\\
        }
        $lms_{stn}$.At($i$) $\gets$ Predict late minutes at $crnt_{stn}$ for $df_{stn}$ using $mdl_{2}^{crnt_{stn}}$ model\\
    }
    \Else{\Comment{$crnt_{stn}$ is at position i $\geq$ 3 during the journey}\\
    	$df_{stn}$ $\gets$ Prepare $crnt_{stn}$'s $3$-$prev$-$stn$ row data-frame (Table \ref{my-label}) using Table \ref{tab_rddtj} with \textbf{late\_mins\_Stn\textsubscript{1}} set as $lms_{stn}$.At($i$-$1$), \textbf{late\_mins\_Stn\textsubscript{2}} set as $lms_{stn}$.At($i$-$2$) and \textbf{late\_mins\_Stn\textsubscript{3}} set as $lms_{stn}$.At($i$-$3$)\\
        \If{$crnt_{stn}$ $\notin$ $3ps_{list}$}{ 
            $crnt_{stn}$ $\gets$ Get nearest \textit{Known Station} in $3ps_{list}$ using Algorithm \ref{algo_knn}\\
        }
        $lms_{stn}$.At($i$) $\gets$ Predict late minutes at $crnt_{stn}$ for $df_{stn}$ using $mdl_{3}^{crnt_{stn}}$ model\\
    }
}
\end{algorithm}


\subsection{Example} \label{exm1}
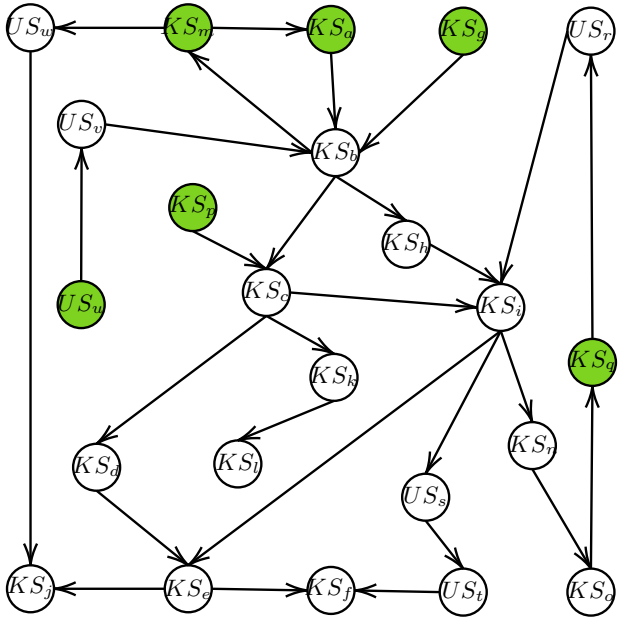
\begin{figure}[b]

  \centering

\tikzset{every picture/.style={line width=0.9pt}} 

\begin{tikzpicture}[x=0.75pt,y=0.75pt,yscale=-1,xscale=1]

\draw    (28.94, 30.34) circle [x radius= 11.85, y radius= 11.85]  ;
\draw  [fill={rgb, 255:red, 126; green, 211; blue, 33 }  ,fill opacity=1 ]  (108.5, 30.34) circle [x radius= 11.85, y radius= 11.85]  ;
\draw  [fill={rgb, 255:red, 126; green, 211; blue, 33 }  ,fill opacity=1 ]  (180.49, 31.18) circle [x radius= 11.85, y radius= 11.85]  ;
\draw  [fill={rgb, 255:red, 126; green, 211; blue, 33 }  ,fill opacity=1 ]  (247.17, 32.02) circle [x radius= 11.85, y radius= 11.85]  ;
\draw    (311.58, 32.02) circle [x radius= 11.85, y radius= 11.85]  ;
\draw  [fill={rgb, 255:red, 255; green, 255; blue, 255 }  ,fill opacity=1 ]  (28.94, 313.47) circle [x radius= 11.85, y radius= 11.85]  ;
\draw    (108.5, 313.47) circle [x radius= 11.85, y radius= 11.85]  ;
\draw  [fill={rgb, 255:red, 255; green, 255; blue, 255 }  ,fill opacity=1 ]  (180.49, 314.31) circle [x radius= 11.85, y radius= 11.85]  ;
\draw    (247.17, 315.15) circle [x radius= 11.85, y radius= 11.85]  ;
\draw  [fill={rgb, 255:red, 255; green, 255; blue, 255 }  ,fill opacity=1 ]  (311.58, 315.15) circle [x radius= 11.85, y radius= 11.85]  ;
\draw    (54.7, 79.07) circle [x radius= 11.85, y radius= 11.85]  ;
\draw  [fill={rgb, 255:red, 126; green, 211; blue, 33 }  ,fill opacity=1 ]  (54.49, 170.05) circle [x radius= 11.85, y radius= 11.85]  ;
\draw    (62.28, 252.14) circle [x radius= 11.85, y radius= 11.85]  ;
\draw    (182.76, 93.35) circle [x radius= 11.85, y radius= 11.85]  ;
\draw  [fill={rgb, 255:red, 126; green, 211; blue, 33 }  ,fill opacity=1 ]  (110.78, 121.07) circle [x radius= 11.85, y radius= 11.85]  ;
\draw    (218.38, 139.56) circle [x radius= 11.85, y radius= 11.85]  ;
\draw    (266.12, 171.48) circle [x radius= 11.85, y radius= 11.85]  ;
\draw    (282.03, 241.21) circle [x radius= 11.85, y radius= 11.85]  ;
\draw    (147.91, 163.92) circle [x radius= 11.85, y radius= 11.85]  ;
\draw    (182, 206.77) circle [x radius= 11.85, y radius= 11.85]  ;
\draw  [fill={rgb, 255:red, 255; green, 255; blue, 255 }  ,fill opacity=1 ]  (133.51, 250.45) circle [x radius= 11.85, y radius= 11.85]  ;
\draw    (228.23, 267.26) circle [x radius= 11.85, y radius= 11.85]  ;
\draw  [fill={rgb, 255:red, 126; green, 211; blue, 33 }  ,fill opacity=1 ]  (312.34, 199.21) circle [x radius= 11.85, y radius= 11.85]  ;
\draw [color={rgb, 255:red, 0; green, 0; blue, 0 }  ,draw opacity=1 ]   (180.49,43.03) -- (182.76,81.49) ;
\draw [shift={(182.76,81.49)}, rotate = 266.62] [color={rgb, 255:red, 0; green, 0; blue, 0 }  ,draw opacity=1 ]   (0,0) .. controls (3.31,-0.3) and (6.95,-1.4) .. (10.93,-3.29)(0,0) .. controls (3.31,0.3) and (6.95,1.4) .. (10.93,3.29)   ;

\draw [color={rgb, 255:red, 0; green, 0; blue, 0 }  ,draw opacity=1 ]   (182.76,105.2) -- (147.91,152.07) ;
\draw [shift={(147.91,152.07)}, rotate = 306.64] [color={rgb, 255:red, 0; green, 0; blue, 0 }  ,draw opacity=1 ]   (0,0) .. controls (3.31,-0.3) and (6.95,-1.4) .. (10.93,-3.29)(0,0) .. controls (3.31,0.3) and (6.95,1.4) .. (10.93,3.29)   ;

\draw [color={rgb, 255:red, 0; green, 0; blue, 0 }  ,draw opacity=1 ]   (147.91,175.77) -- (62.28,240.28) ;
\draw [shift={(62.28,240.28)}, rotate = 323.01] [color={rgb, 255:red, 0; green, 0; blue, 0 }  ,draw opacity=1 ]   (0,0) .. controls (3.31,-0.3) and (6.95,-1.4) .. (10.93,-3.29)(0,0) .. controls (3.31,0.3) and (6.95,1.4) .. (10.93,3.29)   ;

\draw [color={rgb, 255:red, 0; green, 0; blue, 0 }  ,draw opacity=1 ]   (62.28,263.99) -- (108.5,301.61) ;
\draw [shift={(108.5,301.61)}, rotate = 219.14] [color={rgb, 255:red, 0; green, 0; blue, 0 }  ,draw opacity=1 ]   (0,0) .. controls (3.31,-0.3) and (6.95,-1.4) .. (10.93,-3.29)(0,0) .. controls (3.31,0.3) and (6.95,1.4) .. (10.93,3.29)   ;

\draw [color={rgb, 255:red, 0; green, 0; blue, 0 }  ,draw opacity=1 ]   (120.36,313.47) -- (168.64,314.31) ;
\draw [shift={(168.64,314.31)}, rotate = 181] [color={rgb, 255:red, 0; green, 0; blue, 0 }  ,draw opacity=1 ]   (0,0) .. controls (3.31,-0.3) and (6.95,-1.4) .. (10.93,-3.29)(0,0) .. controls (3.31,0.3) and (6.95,1.4) .. (10.93,3.29)   ;

\draw [color={rgb, 255:red, 0; green, 0; blue, 0 }  ,draw opacity=1 ]   (247.17,43.87) -- (194.62,93.35) ;
\draw [shift={(194.62,93.35)}, rotate = 316.73] [color={rgb, 255:red, 0; green, 0; blue, 0 }  ,draw opacity=1 ]   (0,0) .. controls (3.31,-0.3) and (6.95,-1.4) .. (10.93,-3.29)(0,0) .. controls (3.31,0.3) and (6.95,1.4) .. (10.93,3.29)   ;

\draw [color={rgb, 255:red, 0; green, 0; blue, 0 }  ,draw opacity=1 ]   (182.76,105.2) -- (218.38,127.7) ;
\draw [shift={(218.38,127.7)}, rotate = 212.28] [color={rgb, 255:red, 0; green, 0; blue, 0 }  ,draw opacity=1 ]   (0,0) .. controls (3.31,-0.3) and (6.95,-1.4) .. (10.93,-3.29)(0,0) .. controls (3.31,0.3) and (6.95,1.4) .. (10.93,3.29)   ;

\draw [color={rgb, 255:red, 0; green, 0; blue, 0 }  ,draw opacity=1 ]   (230.23,139.56) -- (266.12,159.63) ;
\draw [shift={(266.12,159.63)}, rotate = 209.22] [color={rgb, 255:red, 0; green, 0; blue, 0 }  ,draw opacity=1 ]   (0,0) .. controls (3.31,-0.3) and (6.95,-1.4) .. (10.93,-3.29)(0,0) .. controls (3.31,0.3) and (6.95,1.4) .. (10.93,3.29)   ;

\draw [color={rgb, 255:red, 0; green, 0; blue, 0 }  ,draw opacity=1 ]   (266.12,183.34) -- (108.5,301.61) ;
\draw [shift={(108.5,301.61)}, rotate = 323.11] [color={rgb, 255:red, 0; green, 0; blue, 0 }  ,draw opacity=1 ]   (0,0) .. controls (3.31,-0.3) and (6.95,-1.4) .. (10.93,-3.29)(0,0) .. controls (3.31,0.3) and (6.95,1.4) .. (10.93,3.29)   ;

\draw [color={rgb, 255:red, 0; green, 0; blue, 0 }  ,draw opacity=1 ]   (96.65,313.47) -- (40.79,313.47) ;
\draw [shift={(40.79,313.47)}, rotate = 360] [color={rgb, 255:red, 0; green, 0; blue, 0 }  ,draw opacity=1 ]   (0,0) .. controls (3.31,-0.3) and (6.95,-1.4) .. (10.93,-3.29)(0,0) .. controls (3.31,0.3) and (6.95,1.4) .. (10.93,3.29)   ;

\draw [color={rgb, 255:red, 0; green, 0; blue, 0 }  ,draw opacity=1 ]   (120.36,30.34) -- (168.64,31.18) ;
\draw [shift={(168.64,31.18)}, rotate = 181] [color={rgb, 255:red, 0; green, 0; blue, 0 }  ,draw opacity=1 ]   (0,0) .. controls (3.31,-0.3) and (6.95,-1.4) .. (10.93,-3.29)(0,0) .. controls (3.31,0.3) and (6.95,1.4) .. (10.93,3.29)   ;

\draw    (147.91,175.77) -- (182,194.91) ;
\draw [shift={(182,194.91)}, rotate = 209.3] [color={rgb, 255:red, 0; green, 0; blue, 0 }  ]   (0,0) .. controls (3.31,-0.3) and (6.95,-1.4) .. (10.93,-3.29)(0,0) .. controls (3.31,0.3) and (6.95,1.4) .. (10.93,3.29)   ;

\draw [color={rgb, 255:red, 0; green, 0; blue, 0 }  ,draw opacity=1 ]   (182,218.62) -- (133.51,238.6) ;
\draw [shift={(133.51,238.6)}, rotate = 337.61] [color={rgb, 255:red, 0; green, 0; blue, 0 }  ,draw opacity=1 ]   (0,0) .. controls (3.31,-0.3) and (6.95,-1.4) .. (10.93,-3.29)(0,0) .. controls (3.31,0.3) and (6.95,1.4) .. (10.93,3.29)   ;

\draw [color={rgb, 255:red, 0; green, 0; blue, 0 }  ,draw opacity=1 ]   (266.12,183.34) -- (282.03,229.36) ;
\draw [shift={(282.03,229.36)}, rotate = 250.93] [color={rgb, 255:red, 0; green, 0; blue, 0 }  ,draw opacity=1 ]   (0,0) .. controls (3.31,-0.3) and (6.95,-1.4) .. (10.93,-3.29)(0,0) .. controls (3.31,0.3) and (6.95,1.4) .. (10.93,3.29)   ;

\draw    (282.03,253.07) -- (311.58,303.29) ;
\draw [shift={(311.58,303.29)}, rotate = 239.53] [color={rgb, 255:red, 0; green, 0; blue, 0 }  ]   (0,0) .. controls (3.31,-0.3) and (6.95,-1.4) .. (10.93,-3.29)(0,0) .. controls (3.31,0.3) and (6.95,1.4) .. (10.93,3.29)   ;

\draw [color={rgb, 255:red, 0; green, 0; blue, 0 }  ,draw opacity=1 ]   (110.78,132.93) -- (147.91,152.07) ;
\draw [shift={(147.91,152.07)}, rotate = 207.27] [color={rgb, 255:red, 0; green, 0; blue, 0 }  ,draw opacity=1 ]   (0,0) .. controls (3.31,-0.3) and (6.95,-1.4) .. (10.93,-3.29)(0,0) .. controls (3.31,0.3) and (6.95,1.4) .. (10.93,3.29)   ;

\draw [color={rgb, 255:red, 0; green, 0; blue, 0 }  ,draw opacity=1 ]   (159.76,163.92) -- (254.26,171.48) ;
\draw [shift={(254.26,171.48)}, rotate = 184.57] [color={rgb, 255:red, 0; green, 0; blue, 0 }  ,draw opacity=1 ]   (0,0) .. controls (3.31,-0.3) and (6.95,-1.4) .. (10.93,-3.29)(0,0) .. controls (3.31,0.3) and (6.95,1.4) .. (10.93,3.29)   ;

\draw [color={rgb, 255:red, 0; green, 0; blue, 0 }  ,draw opacity=1 ]   (311.58,303.29) -- (312.34,211.06) ;
\draw [shift={(312.34,211.06)}, rotate = 450.47] [color={rgb, 255:red, 0; green, 0; blue, 0 }  ,draw opacity=1 ]   (0,0) .. controls (3.31,-0.3) and (6.95,-1.4) .. (10.93,-3.29)(0,0) .. controls (3.31,0.3) and (6.95,1.4) .. (10.93,3.29)   ;

\draw [color={rgb, 255:red, 0; green, 0; blue, 0 }  ,draw opacity=1 ]   (312.34,187.35) -- (311.58,43.87) ;
\draw [shift={(311.58,43.87)}, rotate = 449.7] [color={rgb, 255:red, 0; green, 0; blue, 0 }  ,draw opacity=1 ]   (0,0) .. controls (3.31,-0.3) and (6.95,-1.4) .. (10.93,-3.29)(0,0) .. controls (3.31,0.3) and (6.95,1.4) .. (10.93,3.29)   ;

\draw [color={rgb, 255:red, 0; green, 0; blue, 0 }  ,draw opacity=1 ]   (299.73,32.02) -- (266.12,159.63) ;
\draw [shift={(266.12,159.63)}, rotate = 284.76] [color={rgb, 255:red, 0; green, 0; blue, 0 }  ,draw opacity=1 ]   (0,0) .. controls (3.31,-0.3) and (6.95,-1.4) .. (10.93,-3.29)(0,0) .. controls (3.31,0.3) and (6.95,1.4) .. (10.93,3.29)   ;

\draw [color={rgb, 255:red, 0; green, 0; blue, 0 }  ,draw opacity=1 ]   (266.12,183.34) -- (228.23,255.4) ;
\draw [shift={(228.23,255.4)}, rotate = 297.73] [color={rgb, 255:red, 0; green, 0; blue, 0 }  ,draw opacity=1 ]   (0,0) .. controls (3.31,-0.3) and (6.95,-1.4) .. (10.93,-3.29)(0,0) .. controls (3.31,0.3) and (6.95,1.4) .. (10.93,3.29)   ;

\draw [color={rgb, 255:red, 0; green, 0; blue, 0 }  ,draw opacity=1 ]   (228.23,279.11) -- (247.17,303.29) ;
\draw [shift={(247.17,303.29)}, rotate = 231.92000000000002] [color={rgb, 255:red, 0; green, 0; blue, 0 }  ,draw opacity=1 ]   (0,0) .. controls (3.31,-0.3) and (6.95,-1.4) .. (10.93,-3.29)(0,0) .. controls (3.31,0.3) and (6.95,1.4) .. (10.93,3.29)   ;

\draw [color={rgb, 255:red, 0; green, 0; blue, 0 }  ,draw opacity=1 ]   (235.32,315.15) -- (192.34,314.31) ;
\draw [shift={(192.34,314.31)}, rotate = 361.12] [color={rgb, 255:red, 0; green, 0; blue, 0 }  ,draw opacity=1 ]   (0,0) .. controls (3.31,-0.3) and (6.95,-1.4) .. (10.93,-3.29)(0,0) .. controls (3.31,0.3) and (6.95,1.4) .. (10.93,3.29)   ;

\draw [color={rgb, 255:red, 0; green, 0; blue, 0 }  ,draw opacity=1 ]   (66.56,79.07) -- (170.91,93.35) ;
\draw [shift={(170.91,93.35)}, rotate = 187.79] [color={rgb, 255:red, 0; green, 0; blue, 0 }  ,draw opacity=1 ]   (0,0) .. controls (3.31,-0.3) and (6.95,-1.4) .. (10.93,-3.29)(0,0) .. controls (3.31,0.3) and (6.95,1.4) .. (10.93,3.29)   ;

\draw [color={rgb, 255:red, 0; green, 0; blue, 0 }  ,draw opacity=1 ]   (170.91,93.35) -- (108.5,42.19) ;
\draw [shift={(108.5,42.19)}, rotate = 399.34000000000003] [color={rgb, 255:red, 0; green, 0; blue, 0 }  ,draw opacity=1 ]   (0,0) .. controls (3.31,-0.3) and (6.95,-1.4) .. (10.93,-3.29)(0,0) .. controls (3.31,0.3) and (6.95,1.4) .. (10.93,3.29)   ;

\draw [color={rgb, 255:red, 0; green, 0; blue, 0 }  ,draw opacity=1 ]   (54.49,158.2) -- (54.7,90.92) ;
\draw [shift={(54.7,90.92)}, rotate = 450.18] [color={rgb, 255:red, 0; green, 0; blue, 0 }  ,draw opacity=1 ]   (0,0) .. controls (3.31,-0.3) and (6.95,-1.4) .. (10.93,-3.29)(0,0) .. controls (3.31,0.3) and (6.95,1.4) .. (10.93,3.29)   ;

\draw [color={rgb, 255:red, 0; green, 0; blue, 0 }  ,draw opacity=1 ]   (96.65,30.34) -- (40.79,30.34) ;
\draw [shift={(40.79,30.34)}, rotate = 360] [color={rgb, 255:red, 0; green, 0; blue, 0 }  ,draw opacity=1 ]   (0,0) .. controls (3.31,-0.3) and (6.95,-1.4) .. (10.93,-3.29)(0,0) .. controls (3.31,0.3) and (6.95,1.4) .. (10.93,3.29)   ;

\draw [color={rgb, 255:red, 0; green, 0; blue, 0 }  ,draw opacity=1 ]   (28.94,42.19) -- (28.94,301.61) ;
\draw [shift={(28.94,301.61)}, rotate = 270] [color={rgb, 255:red, 0; green, 0; blue, 0 }  ,draw opacity=1 ]   (0,0) .. controls (3.31,-0.3) and (6.95,-1.4) .. (10.93,-3.29)(0,0) .. controls (3.31,0.3) and (6.95,1.4) .. (10.93,3.29)   ;

\draw (28.94,29.92) node  [align=left] {{\small $US_w$}};
\draw (108.5,29.92) node  [align=left] {{\small $KS_m$}};
\draw (180.49,30.76) node  [align=left] {{\small $KS_a$}};
\draw (247.17,31.6) node  [align=left] {{\small $KS_g$}};
\draw (311.58,31.6) node  [align=left] {{\small $US_r$}};
\draw (28.94,313.05) node  [align=left] {{\small $KS_j$}};
\draw (108.5,313.89) node  [align=left] {{\small $KS_e$}};
\draw (179.73,313.89) node  [align=left] {{\small $KS_f$}};
\draw (247.17,314.73) node  [align=left] {{\small $US_t$}};
\draw (311.58,314.73) node  [align=left] {{\small $KS_o$}};
\draw (54.7,78.65) node  [align=left] {{\small $US_v$}};
\draw (54.49,169.63) node  [align=left] {{\small $US_u$}};
\draw (62.28,251.72) node  [align=left] {{\small $KS_d$}};
\draw (182.76,92.93) node  [align=left] {{\small $KS_b$}};
\draw (110.78,120.65) node  [align=left] {{\small $KS_p$}};
\draw (218.38,139.14) node  [align=left] {{\small $KS_h$}};
\draw (266.12,171.06) node  [align=left] {{\small $KS_i$}};
\draw (282.79,240.79) node  [align=left] {{\small $KS_n$}};
\draw (147.91,163.5) node  [align=left] {{\small $KS_c$}};
\draw (182,206.35) node  [align=left] {{\small $KS_k$}};
\draw (133.51,250.03) node  [align=left] {{\small $KS_l$}};
\draw (228.23,266.84) node  [align=left] {{\small $US_s$}};
\draw (312.34,198.79) node  [align=left] {{\small $KS_q$}};

\end{tikzpicture}

  \caption{Visual view of example trains routes $KT_i$ and $UT_j$. Starting stations are highlighted.}
  \label{tr_route}
\end{figure}

In our example, let there be five \textit{Known Trains} ($KT_i$)  routes and two \textit{Unknown Trains} ($UT_j$) routes with dummy stations $KS_\alpha$ and $US_\beta$ to explain our proposed framework, where $KS_\alpha$ $\forall$ $\alpha$ $\in$ (a..q) and $US_\beta$ $\forall$ $\beta$ $\in$ (r..w) are \textit{Known Stations} and \textit{Unknown Stations}, respectively. Figure \ref{tr_route} shows the train route map where source stations are colored green.

\begin{itemize}
\item $KT_1$ Journey: 
$KS_a \rightarrow KS_b \rightarrow KS_c \rightarrow KS_d \rightarrow KS_e \rightarrow KS_f$
\item $KT_2$  Journey: 
$KS_g \rightarrow KS_b \rightarrow KS_h \rightarrow KS_i \rightarrow KS_e \rightarrow KS_j$
\item $KT_3$ Journey: 
$KS_m \rightarrow KS_a \rightarrow KS_b \rightarrow KS_c \rightarrow KS_k \rightarrow KS_l$
\item $KT_4$ Journey: 
$KS_g \rightarrow KS_b \rightarrow KS_h \rightarrow KS_i \rightarrow KS_n \rightarrow KS_o$
\item $KT_5$ Journey: 
$KS_p \rightarrow KS_c \rightarrow KS_i \rightarrow KS_n \rightarrow KS_o \rightarrow KS_q$
\\
\item $UT_1$ Journey: 
$KS_q \rightarrow US_r \rightarrow KS_i \rightarrow US_s \rightarrow US_t \rightarrow KS_f$
\item $UT_2$ Journey: 
$US_u \rightarrow US_v \rightarrow KS_b \rightarrow KS_m \rightarrow US_w \rightarrow KS_j$
\end{itemize}


\begin{algorithm}[]
\SetAlgoLined
\small
\caption{$k$-NN search framework to get a \textit{Known Station} best similar  to any type of Station ($k$ set to 10)}
\label{algo_knn}
\KwIn{A Station $stn_{S}$, Valid $ips_{list}$ of \textit{Known Stations}}
\KwOut{A nearest \textit{Known Station} $stn_{KS}$}
$stn_{KS}^{nll}$ $\gets$ Get $k$-NN \textit{Known Stations} to $stn_{S}$ among stations in $ips_{list}$ on the basis of Latitude and Longitude \\
$stn_{KS}^{ndt}$ $\gets$ Get $k$-NN \textit{Known Stations} to $stn_{S}$ among stations in $stn_{KS}^{nll}$ on the basis of Degree and Traffic \\
Return the first station among $stn_{KS}^{ndt}$\\
\end{algorithm}

\subsubsection{Data Preparation and Training} \label{dcp}
We collect \textit{Train Data} Table \ref{tab_rddtj} for each of the seven trains and divide them into two categories: \textit{Known Trains} ($KT_i$ $\forall$ $i$ $\in$ $<1..5>$) and \textit{Unknown Trains} ($UT_j$ $\forall$ $j$ $\in$ $<1..2>$) based on the amount of data collected for each train. After the actual segregation of collected data as showed in \figurename \ref{pict_rep_data}, we 
prepare $n$-$prev$-$stn$ data-frame Table \ref{my-label} for each $KS_\alpha$ using $KT_i$'s Table \ref{tab_rddtj} data.

\begin{itemize}
\item Preparation of $n$-$prev$-$stn$ data-frames Table \ref{my-label} for $KS_a$:\\
We prepare a $1$-$prev$-$stn$ data-frame for $KS_a$ owing to Train $KT_3$ only since it has $KS_m$ as one  station previous to it. It is navigated by $KT_1$ also, but it is the source station there, thus has zero stations previous to it.
\item Preparation of $n$-$prev$-$stn$ data-frames Table \ref{my-label} for $KS_b$:\\
We prepare a $1$-$prev$-$stn$ data-frame for $KS_b$ owing to trains $KT_1$, $KT_2$, $KT_3$ and $KT_4$ since it has a valid set of one station previous to it and a $2$-$prev$-$stn$ data-frame owing to train $KT_3$, as it has two stations previous to it.


\item Preparation of $n$-$prev$-$stn$ data-frames Table \ref{my-label} for $KS_c$:\\
We prepare a $1$-$prev$-$stn$ data-frame for it owing to Train $KT_1$, $KT_3$ and $KT_5$ as they have a valid one station previous to $KS_c$ during the journey. Another $2$-$prev$-$stn$ data-frame is prepared for it owing to Train $KT_1$ and $KT_3$, and a $3$-$prev$-$stn$ data-frame owing to Train $KT_3$. 
\end{itemize}

Similarly, for each of the \textit{Known Stations}, we prepare valid $n$-$prev$-$stn$ data-frames Table \ref{my-label}, depending on the number of stations previous to them during the journey of \textit{Known Trains}. Later we use those $n$-$prev$-$stn$ data-frames to train $n$-OMPR models (RFR and RR) for each \textit{Known Station} as explained in Algorithm \ref{algo_training}. 
While training the models, we also maintain a list of stations $ips_{list}$ which stores the names of stations (station codes) which have $i^{th}$-OMPR models. For example, in context of all five \textit{Known Trains} here, the stations in $1ps_{list}$ are ($KS_b$, $KS_c$, $KS_d$, $KS_e$, $KS_f$, $KS_h$, $KS_i$, $KS_j$, $KS_a$, $KS_k$, $KS_l$, $KS_n$, $KS_o$, $KS_q$) since they have one valid station previous to them during the journey of various $KT_i$; ... 
$4ps_{list}$ has stations ($KS_e$, $KS_f$, $KS_j$, $KS_k$, $KS_l$, $KS_n$, $KS_o$, $KS_q$) since each of them has a valid set of $4$ stations previous to them. 

\vspace{5pt}
\subsubsection{Prediction of Late Minutes for Train Journeys}
We explain $N$-OMLMPF algorithm (Algorithm \ref{algo_lmsp}) here with the help of above train examples. We employ a feed-forward method for late minutes prediction at each of the in-line stations where the late minutes predicted for the $n$ previous stations and their other details are incorporated in current target station's $n$-$prev$-$stn$ row data-frame. 
(A row data-frame consists of only one row of Table \ref{my-label}).

\paragraph{Known Trains Late Minutes Prediction} \label{ktlmp}
Stations in-line during the journeys of cross-validation set and the test set of \textit{Known Trains} consist of only \textit{Known Stations} for which we have trained models saved from Algorithm \ref{algo_training}. The column entries in $n$-$prev$-$stn$ row data-frame (Table \ref{my-label}) for the current station at which late minutes are to be predicted are filled accordingly as explained in the table, except \textbf{Stn\textsubscript{0}\_late\_minutes} since we aim to predict it here. 
 Say for train $KT_3$'s cross-validation or test data, we predict late minutes at each station. As per the execution steps of Algorithm \ref{algo_lmsp} the late minutes at: 
\begin{itemize}
\item $KS_m$ is assumed to be $0$ since it is a source station thus list $lms_{stn}$ is $<0>$.
\item $KS_a$ is predicted through $mdl_{1}^{KS_a}$ since we have this $1$-OMPR model trained over the $1$-$prev$-$stn$ training data-frame for $KS_a$. We fill the $1$-$prev$-$stn$ row data-frame for $KS_a$ with $Stn_1$ set as $KS_m$ and late minutes at $Stn_1$ set as the first entry in $lms_{stn}$ i.e. $0$. Say the predicted late minutes at $KS_a$ is 5, hence $lms_{stn}$ extends to $<0, 5>$.

\item $KS_b$ is predicted through $mdl_{2}^{KS_b}$ as we have this $2$-OMPR model trained for it. 
 The first and second entry in $lms_{stn}$ list, ($0$ and $5$) are used as late minutes at station $Stn_2$ and $Stn_1$ respectively in the $2$-$prev$-$stn$ row data-frame for station $KS_b$ to predict the late minutes at it; say $10$ minutes. So the list $lms_{stn}$ becomes $<0, 5, 10>$.
\item In a similar fashion, we keep feed-forwarding the predicted late minutes at previous stations to predict the late minutes at $KS_c$, $KS_k$, and $KS_l$ through $3$-OMPR models $mdl_{3}^{KS_c}$, $mdl_{3}^{KS_k}$, and $mdl_{3}^{KS_l}$ respectively. 
\end{itemize}
\paragraph{Unknown Trains Late Minutes Prediction}
We choose train $UT_2$ for explaining Algorithm \ref{algo_lmsp} to predict late minutes for \textit{Unknown Trains}' in-line stations. The late minutes at: 


\begin{itemize}
\item $US_u$ is assumed to be $0$ since it is the source station. Thus the late minutes list $lms_{stn}$ is initialized with $<0>$.

\item $US_v$ is predicted as follows. We do not have a trained $1$-OMPR model (neither RFR nor RR) for $US_v$ since it is an \textit{Unknown Station}, thus not in $1ps_{list}$. Hence, via Algorithm \ref{algo_knn} we find a \textit{Known Station} nearest to it among the ones in $1ps_{list}$ which have a $1$-OMPR model (RFR and RR), say station $KS_a$ is found. 
Next, the $1$-$prev$-$stn$ row data-frame prepared for $US_v$ with $US_u$ set as $Stn_1$ is fed to the model $mdl_{1}^{KS_a}$ to predict late minutes at $US_v$, say $10$ minutes. Thus $lms_{stn}$ list extends to $<0, 10>$.

\item $KS_b$ is predicted through model $mdl_{2}^{KS_b}$ with $Stn_1$, $Stn_2$ and late minutes at $Stn_1$, late minutes at $Stn_2$ set as $US_v$, $US_u$ and $10$, $0$ respectively; say $15$ minutes is predicted, thus the list $lms_{stn}$ becomes $<0, 10, 15>$. 

\item $KS_m$ is predicted as follows. 
It can be noticed from above set of \textit{Known Trains} journey that we do not have a valid trained model $mdl_{3}^{KS_m}$ in spite of the current target station being a \textit{Known Station} since no $3$-$prev$-$stn$ data-frame for station $KS_m$ could be prepared from any of the \textit{Known Trains}. So we choose a station among $3ps_{list}$ which is best similar to $KS_m$ through Algorithm \ref{algo_knn} (say station $KS_e$ is chosen). Thus $mdl_{3}^{KS_e}$ is used to predict the late minutes (say $40$ minutes) on the row data-frame for $KS_m$ with $Stn_1$, $Stn_2$, and $Stn_3$ being $KS_b$, $US_v$, and $US_u$ respectively with corresponding late minutes as $15$, $10$ and $0$. Thus the list becomes $<0, 10, 15, 40>$.

\item $US_w$ is predicted through a $3$-OMPR model; say $mdl_{3}^{KS_i}$ where $KS_i$ is obtained through Algorithm \ref{algo_knn} for $US_w$. The $3$-$prev$-$stn$ row data-frame for it has $KS_m$, $KS_b$, $US_v$ set as $Stn_1$, $Stn_2$, and $Stn_3$ respectively. 


\item $KS_j$ is predicted through model $mdl_{3}^{KS_j}$ on its $3$-$prev$-$stn$ row data-frame with $US_w$, $KS_m$, and $KS_b$ set as $Stn_1$, $Stn_2$, and $Stn_3$ respectively.

\end{itemize}
\section{Experiments and Result Analysis} \label{ra}
The $N$-OMLMPF Algorithm \ref{algo_lmsp} was executed on three sets of data, namely Cross-validation Data of \textit{Known Trains}, Test Data of \textit{Known Trains} and Test Data of  \textit{Unknown Trains} as mentioned in Figure \ref{pict_rep_data} for different values of $N$ (in $N$-OMLMPF). 
We enumerate four detailed experiments below, which were conducted with both RFR and RR models individually:

\begin{enumerate}
\item Exp 1: We ignored {\textbf{tfc\_of\_Stn\textsubscript{i}}}, {\textbf{deg\_of\_Stn\textsubscript{i}}} and \textbf{Stn\textsubscript{i}\_dfs} columns from data-frame Table \ref{my-label} since these features are implicitly captured in \textbf{Stn\textsubscript{i}\_code}. Experiment was conducted on dataset \textit{52TrnsTrCv}.

\item Exp 2: We ignored the \textbf{Stn\textsubscript{i}\_code} columns from data-frame Table \ref{my-label} as {\textbf{tfc\_of\_Stn\textsubscript{i}}}, {\textbf{deg\_of\_Stn\textsubscript{i}}} and \textbf{Stn\textsubscript{i}\_dfs} numerically capture the property of station codes. This was done for \textit{Unknown Trains} case because we did not have partial consecutive in-line station path of $KS$s and $US$s (hence no \textbf{Stn\textsubscript{i}\_code}s) due to the test data being Zero-Shot. The experiment was conducted on \textit{83TrnsTe} data after learning the prediction models from \textit{52TrnsTrCv} data to assess the transfer of knowledge from \textit{Known Trains} to \textit{Unknown Trains}. 

\item Exp 3: We conducted Exp 2 again on \textit{52TrnsTrCv} data, where results similar to that obtained in Exp 1 for cross-validation data endorses our notion of vice-versa representation of stations, as done in Exp 1 and Exp 2.

\item Exp 4: We conducted Exp 2 on \textit{52TrnsTe} data with prediction models learned from \textit{52TrnsTrCv} data.
\end{enumerate}

After conducting the experiments 
we analyzed the results 
to evaluate the performance of trained models and to determine the optimum value of $N$ (in $N$-OMLMPF). 
For brevity, we do not present the detailed results for all 135 trains, but we do justice by presenting $4$-OMLMPF output on test data of few trains in Tables \ref{Tr22811}, \ref{Tr12326}, \ref{Tr12356} (negative numbers in tables suggest that the train arrived early by those many minutes). 

\begin{table*}[h]
\centering
\caption{Predicted Late Minutes for \textit{Known Train} 22811 Test Data (obtained from $4$-OMLMPF with RFR models)}
\label{Tr22811}
\begin{tabular}{|c|c|c|c|c|c|c|c|c|c|c|c|c|c|c|}
\hline
\textbf{Stations:}                                                         & \textbf{BBS} & \textbf{CTC} & \textbf{JJKR} & \textbf{BHC} & \textbf{BLS} & \textbf{KGP} & \textbf{BQA} & \textbf{ADRA} & \textbf{GMO} & \textbf{KQR} & \textbf{GAYA} & \textbf{MGS} & \textbf{CNB} & \textbf{NDLS} \\ \hline
\textbf{\begin{tabular}[c]{@{}c@{}}Actual Late\\ Minutes:\end{tabular}}    & 0            & 2            & 8             & -1           & 13           & 25           & 19           & 18            & 2            & 9            & -21           & -5           & 6            & 15            \\ \hline
\textbf{\begin{tabular}[c]{@{}c@{}}Predicted Late\\ Minutes:\end{tabular}} & 0            & 2.75         & 6.83          & 0.01         & 17.44        & 16.52        & 11.22        & 17.65         & 1.94         & 16.01        & -8.77         & -0.25        & 12.26        & 23.10         \\ \hline
\end{tabular}
\vspace{-5pt}
\end{table*}

\begin{table*}[h]

\centering
\setlength\tabcolsep{3pt}
\caption{Predicted Late Minutes for \textit{Known Train} 12326 Test Data (obtained from $4$-OMLMPF with RFR models)}
\label{Tr12326}
\begin{tabular}{|c|c|c|c|c|c|c|c|c|c|c|c|c|c|c|c|c|c|}
\hline
\textbf{Stations:}                                                         & \textbf{NLDM} & \textbf{ANSB} & \textbf{RPAR} & \textbf{SIR} & \textbf{UMB} & \textbf{SRE} & \textbf{MB} & \textbf{BE} & \textbf{LKO} & \textbf{BSB} & \textbf{MGS} & \textbf{PNBE} & \textbf{KIUL} & \textbf{JAJ} & \textbf{JSME} & \textbf{ASN} & \textbf{KOAA} \\ \hline
\textbf{\begin{tabular}[c]{@{}c@{}}Actual Late\\ Minutes:\end{tabular}}    & 0             & 3             & 4             & -11          & 0            & -6           & 15          & 55          & 30           & 10           & 18           & 10            & 11            & 0            & 7             & 3            & 5             \\ \hline
\textbf{\begin{tabular}[c]{@{}c@{}}Predicted Late\\ Minutes:\end{tabular}} & 0             & 9.38          & 7.87          & -2.43        & 3.61         & 0.50         & 26.13       & 36.14       & 29.42        & 32.14        & 20.38        & 3.296         & 6.87          & -3.80        & 17.55         & 14.30        & 13.91         \\ \hline
\end{tabular}
\vspace{-5pt}
\end{table*}

\begin{table*}[h]

\centering
\setlength\tabcolsep{2.5pt}
\caption{Predicted Late Minutes for \textit{Unknown Train} 12356 Test Data with 3 \textit{Unknown Stations} (obtained from $4$-OMLMPF with RFR models)}
\label{Tr12356}
\begin{tabular}{|c|c|c|c|c|c|c|c|c|c|c|c|c|c|c|c|c|c|c|c|}
\hline
\textbf{Stations:}                                                         & \textbf{JAT} & \textbf{PTKC} & \textbf{JRC} & \textbf{LDH} & \textbf{UMB} & \textbf{SRE} & \textbf{MB} & \textbf{BE} & \textbf{LKO} & \textbf{RBL} & \textbf{JAIS} & \textbf{AME} & \textbf{PBH} & \textbf{BOY} & \textbf{BSB} & \textbf{MGS} & \textbf{DNR} & \textbf{PNBE} & \textbf{RJPB} \\ \hline
\textbf{\begin{tabular}[c]{@{}c@{}}Actual Late\\ Minutes:\end{tabular}}    & 0            & 8             & 3            & 0            & -5           & -15          & -10         & -1          & 30           & 41           & 51            & 57           & 74           & 111          & 75           & 123          & 130          & 120           & 120           \\ \hline
\textbf{\begin{tabular}[c]{@{}c@{}}Predicted Late\\ Minutes:\end{tabular}} & 0            & 10.19         & 10.74        & 10.17        & 11.60        & 11.97        & 27.24       & 34.63       & 28.45        & 40.15        & 41.29         & 42.94        & 60.71        & 72.51        & 75.25        & 70.50        & 74.45        & 67.95         & 71.80         \\ \hline
\end{tabular}
\vspace{-5pt}
\end{table*}

\begin{table*}[h]
\setlength\tabcolsep{3.95pt}
\centering
\caption{Confidence Interval (CI) observations for different experiments}
\label{cira}
\begin{tabular}{|c|c|c|c|c|c|c|c|c|c|c|c|c|c|c|c|c|c|c|}
\hline
\multirow{3}{*}{} & \multicolumn{12}{c|}{\textbf{Random Forest Regressor (RFR) Models}}                                                                                                                                   & \multicolumn{6}{c|}{\textbf{Ridge Regressor (RR) Models}}                                         \\ \cline{2-19} 
                  & \multicolumn{3}{c|}{\textbf{Exp 1 (Avg \%age)}} & \multicolumn{3}{c|}{\textbf{Exp 2 (Avg \%age)}} & \multicolumn{3}{c|}{\textbf{Exp 3 (Avg \%age)}} & \multicolumn{3}{c|}{\textbf{Exp 4 (Avg \%age)}} & \multicolumn{3}{c|}{\textbf{Exp 2 (Avg \%age)}} & \multicolumn{3}{c|}{\textbf{Exp 4 (Avg \%age)}} \\ \cline{2-19} 
                  & \textbf{CI68}  & \textbf{CI95}  & \textbf{CI99} & \textbf{CI68}  & \textbf{CI95}  & \textbf{CI99} & \textbf{CI68}  & \textbf{CI95}  & \textbf{CI99} & \textbf{CI68}  & \textbf{CI95}  & \textbf{CI99} & \textbf{CI68}  & \textbf{CI95}  & \textbf{CI99} & \textbf{CI68}  & \textbf{CI95}  & \textbf{CI99} \\ \hline
\textbf{1-OMLMPF} & 34.65          & 61.37          & 70.47         & 5.90           & 14.73          & 18.51         & 33.67          & 61.05          & 70.21         & 27.60          & 55.41          & 65.57         & 4.97           & 12.87          & 17.29         & 22.34          & 44.30          & 55.71         \\ \hline
\textbf{2-OMLMPF} & 35.28          & 61.36          & 70.85         & 5.72           & 14.17          & 18.41         & 33.72          & 61.03          & 70.65         & 27.51          & 56.32          & 66.87         & 5.34           & 12.65          & 16.80         & 22.81          & 43.67          & 56.59         \\ \hline
\textbf{3-OMLMPF} & 33.86          & 62.31          & 71.42         & 6.00           & 14.79          & 18.81         & 33.80          & 62.13          & 71.58         & 27.81          & 55.89          & 66.98         & 4.89           & 12.46          & 16.76         & 22.21          & 44.05          & 55.67         \\ \hline
\textbf{4-OMLMPF} & 34.39          & 62.53          & 71.74         & 5.66           & 14.96          & 18.97         & 33.67          & 61.57          & 71.49         & 27.82          & 55.80          & 66.82         & 4.66           & 12.35          & 16.35         & 21.85          & 43.89          & 55.83         \\ \hline
\textbf{5-OMLMPF} & 34.77          & 62.70          & 72.10         & 5.51           & 14.52          & 18.75         & 33.45          & 62.03          & 71.96         & 27.93          & 56.20          & 67.07         & 4.61           & 12.43          & 16.16         & 21.85          & 43.87          & 55.18         \\ \hline
\end{tabular}
\\~\\
\textbf{CI68}, \textbf{CI95}, and \textbf{CI99} respectively stand for 68\% CI, 95\% CI, and 99\% CI. \textbf{Avg \%age} stands for Average Percentage.
\vspace{-10pt}
\end{table*}

\begin{table*}[]
\centering
\setlength\tabcolsep{1.6pt}
\caption{Mean RMSE values for few \textit{Known Trains} and \textit{Unknown Trains} Test Data (Obtained from 4-OMLMPF with RFR Models)}
\label{tr_rmse}
\begin{tabular}{|l|c|c|c|c|c|c|c|c|c|c|c|c|c|c|c|c|c|c|c|}
\hline
                            & \multicolumn{11}{c|}{\textbf{Known Trains}}                                           & \multicolumn{8}{c|}{\textbf{Unknown Trains}}                  \\ \hline
\textbf{Trains}             & 12305 & 12361 & 12815 & 12307 & 13131 & 13151 & 22811 & 22409 & 18612 & 13119 & 15635 & 03210 & 04401 & 04821 & 12141 & 12295 & 22308 & 12439 & 18311 \\ \hline
\textbf{Number of Journeys} & 16    & 14    & 39    & 84    & 19    & 83    & 28    & 14    & 47    & 25    & 13    & 2     & 1     & 6     & 3     & 4     & 28    & 2     & 3     \\ \hline
\textbf{Mean RMSE}       & 87.12 & 89.38 & 96.61 & 88.26 & 62.84 & 82.34 & 53.71 & 44.72 & 29.42 & 80.66 & 80.22 & 57.37 & 23.86 & 31.97 & 53.38 & 68.49 & 44.83 & 11.75 & 36.20 \\ \hline
\end{tabular}
\\~\\
\textbf{Trains} row consists of unique Train Numbers. \textbf{Number of Journeys} row denotes the number of journeys undertaken by the corresponding train in its Test Data. \textbf{Mean RMSE} row presents the average of the RMSEs of all journeys. For example, Train 12305 covered 16 journeys with a mean RMSE of 87.12.
\vspace{-15pt}
\end{table*}

\vspace{-5pt}
\subsection{Performance Evaluation of Models}
We begin by noting again that a train's \textit{Train Data} consists of multiple instances of journeys, where each journey has the same set of stations that the train plies through. For each in-line station during a train's journey, we calculated monthly 68\%, 95\%, and 99\% Confidence Intervals (CI) around the mean of late minutes in a month, considering the train's complete \textit{Train Data} with outlier late minutes removed by Tukey's Rule \cite{tukey}. For each train's cross-validation/test \textit{Train Data}, the percentage of the number of times the predicted late minutes for an in-line station fell under each matching CI was calculated. Then we averaged out all the percentages (calculated for each train) in different experiments enumerated above. Table \ref{cira} shows the corresponding figures. In Table \ref{tr_rmse} we present the mean Root Mean Square Error (RMSE) values for few \textit{Known Trains} and \textit{Unknown Trains} obtained from their Test Data, where RMSE for a journey was calculated between the predicted late minutes and the actual late minutes. It is to be noted that reported results in Table \ref{cira} and \ref{tr_rmse} are inclusive of journeys where the train actually got late at the source station, but these details could not be captured by our models due to their scarce occurrences.

Preliminary analysis of CI and mean RMSE observations showed that RFR models outperformed RR models. However, for sake of completion, we present CI observations of RR models for some selected experiments in Table \ref{cira}. The scattering of individual late minutes at a station during a month; as observed in Figures \ref{12307}, \ref{Picture10}, and \ref{Picture5} suggests to consider CI95 (or higher) since the late minutes are not closely centered around mean but cover a wider distribution around it. 
Under RFR Models column in Table \ref{cira}, the figures in CI95 columns for Exp 1 and Exp 3 suggest that at an average we were able to predict late minutes at in-line stations during cross-validation journey data of \textit{Known Trains} for approximately 62\% times within 95\% CI (say \textit{accuracy} is 62\%). Figures in Exp 2 under both RFR and RR Models columns in Table \ref{cira} for \textit{Unknown Trains}' test data do not seem promising, but since these results are for Zero-Shot trains for which significant amount of data is not available, 
the observations are appreciable. One should also note here the low mean RMSE values for \textit{Unknown Trains} in Table \ref{tr_rmse}. The higher \textit{accuracies} (around 56\% and 66\% for CI95 and CI99) for \textit{Known Trains}' test data in Exp 4 column under RFR Models column compared to that under RR Models column signify a very important conclusion. Random Forest Regressors (which are an ensemble of multiple decision trees) very well model the deciding factors (in Table \ref{my-label}) compared to  Ridge Regressors, thus the results state that the prediction of late minutes is effectively a decision-based regression task. 

\subsection{Determination of Optimum value of $N$ in $N$-OMLMPF}
We executed Algorithm \ref{algo_lmsp} with values of $N$ $\in$ ($1$..$5$), but which one truly captures the Markov Process property of delays along a train's journey? To answer this we employ two common model selection criterion \cite{aicbic}: Akaike Information Criterion (AIC) and Schwartz Bayesian Information Criterion (BIC) to choose the statistically best regression model.
\begin{equation}
AIC = n\times\text{ln}\Big(\frac{SSE}{n}\Big) + 2p
\end{equation}
\begin{equation}
BIC = n\times\text{ln}\Big(\frac{SSE}{n}\Big) + p\times\text{ln}(n)
\end{equation}
where $n$ stands for the number of observations used to train a model, $SSE$ is the Squared Sum of Errors (between predicted late minutes and the actual late minutes) and $p$ is the number of parameters in the model (number of columns in formatted data-frame Table \ref{my-label}). Lower the score, better the model. 
The count of the number of times a run of $N$-OMLMPF (for a particular value of $N$) yielded the least AIC and BIC scores among all five runs for each train in all four experiments is noted in Table \ref{aicbic}. In Table \ref{aicbic} we see that 
delays along journey undertaken by 40.38\% to 67.30\% of \textit{Known Trains} under related experiments follow a $1$-Order Markov Process since 1-OMLMPF scores minimum AIC and BIC score among other frameworks. Similarly 71.08\% to 81.93\% of \textit{Unknown Trains} follow a $1$-Order Markov Process. Rest of the trains follow a higher order Markov Process with diminishing indications. However lower cumulative RMSE scores (summed over all trains) obtained for $3$- and $4$-OMLMPF under different experimental settings suggest to use them for real-time deployment.



\begin{table}[]
\setlength\tabcolsep{2.4pt}
\centering
\caption{BIC and AIC analysis of $N$-OMLMPF with RFR Models}
\label{aicbic}
\begin{tabular}{|c|c|c|c|c|c|c|c|c|}
\hline
\multicolumn{9}{|c|}{\textbf{Random Forest Regressor Models}}                                                                                                     \\ \hline
\multirow{2}{*}{\textbf{}} & \multicolumn{4}{c|}{\textbf{BIC Analysis}}                        & \multicolumn{4}{c|}{\textbf{AIC Analysis}}                        \\ \cline{2-9} 
                           & \textbf{Exp 1} & \textbf{Exp 2} & \textbf{Exp 3} & \textbf{Exp 4} & \textbf{Exp 1} & \textbf{Exp 2} & \textbf{Exp 3} & \textbf{Exp 4} \\ \hline
\textbf{1-OMLMPF}          & 32             & 68             & 35             & 29             & 21             & 59             & 31             & 23             \\ \hline
\textbf{2-OMLMPF}          & 7              & 7              & 9              & 14             & 9              & 12             & 9              & 10             \\ \hline
\textbf{3-OMLMPF}          & 9              & 5              & 6              & 5              & 12             & 7              & 7              & 11             \\ \hline
\textbf{4-OMLMPF}          & 4              & 3              & 1              & 4              & 8              & 2              & 3              & 6              \\ \hline
\textbf{5-OMLMPF}          & 0              & 0              & 1              & 0              & 2              & 3              & 2              & 2              \\ \hline
\end{tabular}
\\~\\
The figures in each cell denote the number of times an $N$-OMLMPF scored minimum score among other runs, 
e.g. in \textbf{BIC Analysis} column for \textbf{Exp 1}, \textbf{1-OMLMPF} scored minimum BIC score for 32 trains among other runs. 
\vspace{-10pt}
\end{table}

\section{Conclusion and Future Work} \label{c}
Our objective was to predict the late minutes at an in-line station given the route information of a train and a valid date. The significant \textit{accuracy} results in Table \ref{cira} for \textit{Known Trains}' and \textit{Unknown Trains}' data demonstrates the efficacy of our proposed algorithm 
for a highly dynamic problem. We also determine experimentally and statistically that the delays along journey for most of the  trains follow a $1$-Order  Markovian Process, while other few trains follow a higher order Markovian Process. Reasonably low RMSE results obtained for \textit{Unknown Trains} in Table \ref{tr_rmse} also show that we were able to transfer knowledge from \textit{Known Trains} to \textit{Unknown Trains}. The $N$-OMLMPF algorithm is so designed that it can leverage different types of prediction models and predict delay at stations for any train, thus it is train-agnostic. With just $1.2$\% of total trains in India, our approach was able to cover more than $11.3$\% of stations, thereby illustrating scalability . There are many avenues for future work:
(a) one can expand the data collection and extend the analysis to trains India-wide, (b) one can also explore other approaches like time series prediction and neural networks. 
In particular, Recurrent Neural Networks (RNN) have
the property of memorizing past details and predicting the next state. 
The prediction of delays along stations is inherently dynamic  which implicitly calls for an online learning algorithm to continuously learn the changing behavior of railway network and delays. Thus one can attempt to develop an Online RNN algorithm for it. One can also consider predicting delay of trains in other countries.
\vspace{-10pt}
\section{Acknowledgment}
We would like to thank Debarun Bhattacharjya for his help in statistically discovering the  order of Markovian delays through mathematical equations. We also thank Nutanix Technologies India Pvt Ltd for the computational resources. 
\\

\bibliographystyle{IEEEtranS.bst}
\bibliography{references}


\end{document}